\documentclass{amsart}
\usepackage[hidelinks]{hyperref}
\usepackage{multicol}
\usepackage{amsmath}
\usepackage{amssymb}
\usepackage{subfig}
\usepackage{mathtools}
\usepackage{nicematrix,tikz}

\newcommand\ddfrac[2]{\frac{\displaystyle #1}{\displaystyle #2}}

\theoremstyle{definition}

\theoremstyle{remark}

\numberwithin{equation}{section}

\begin{document}
\pagestyle{plain}
\title{Conceptualizing the chaotic perception, demands and challenges}

\author{Amir M. Majd}
\address{Faculty of Physics, University of Tabriz, Tabriz, Iran}
\email{amir.mahmoudimajd@gmail.com}




\begin{abstract}
One propounded theory for the presence of chaos in biological neural networks is that it could be involved in discriminating different olfactory stimuli. Inspired by the idea, in this paper, we define the visual ``chaotic perception'' and spell out the challenges we face when conceptualizing it.
\end{abstract}

\maketitle
\section{Defining the chaotic perception}
Topologically equivalent dynamical systems\cite{Kuznetsov:2004} and linking matrix of strange attractors\cite{Gilmore:2011} are what we need to characterize the act of recognition chaotically. Let us review topological equivalence in the first place. \\
\subsection{Topological equivalence} Assume that $S$ is the state space of a dynamical system, and $\varphi^{t}:\,S\to S,t\in \tau$, $\tau$ being a number set, is the evolution operator, which satisfies $\varphi^{0}X=X,\varphi^{t+t'}=\varphi^{t}(\varphi^{t'})$. An ``orbit'' (or trajectory) setting off from $X_{0}\in S$ is an ordered subset of the state space, i.e, $\{X\in S;\,X=\varphi^{t}X_{0},t\in \tau\}$. \textit{Two dynamical systems are topologically equivalent if there is a homeomorphism $\xi:S_{1}\to S_{2}$ mapping orbits of the first system onto the second one.}\\
If $S\subset X$, for any $t\in \tau$ $\varphi^{t}S\subseteq S$ is called an invariant set. Clearly, each orbit is an invariant set. Invariant subsets are stable if for any $U\supset S$ ($U$ a small neighborhood), there exists $V$ (another neighborhood) such that  $\varphi^{t}x \in U$ for any $x\in V,t\in T$. Or, there exists $U\supset S$ for any $x\in U$ we have $\varphi^{t}x\to S$ as $t\to \infty$. The former one is called Lyapunov stability and the latter one is said to be asymptotic stability\footnote{None of the aforementioned stabilities implies the other one.}. Seeing such stable invariant sets (or attractors) as geometric objects, they are termed strange if their dimension is fractional.\\
\subsection{Linking matrix of the strange attractors} There are two sets of attributes of strange attractors, dynamical and topological. The former one, known as ``old'' chaos quantifiers, are fractal dimension, the metric entropy, and the spectrum of Lyapunov exponents\cite{Hao:1989}. The latter attribute, all being topological invariants (or fingerprints) of strange attractors, could be linking numbers, topological entropy, and rotation rates (if the associated phase space is $\mathbb{R}^{3}\times S^{1}$). Because the topological organization breaks down in higher dimensions, the finger prints are applicable merely in 3-space\cite{Gilmore:2011}. Let us focus on the first topological property.

In a chaotic time series, there are sections that are so close to some periodic orbits. These sections are called ``surrogate periodic orbits'', the stabilization of which yields Unstable Periodic Orbits (UPOs). Then, the motion on the attractor may be conceived as hopping from one UPO to another.
In this paper, we choose to work with linking numbers. In principle, to find the linking matrix superpose the extracted UPOs, detect the crossings, and use the conventions depicted in fig.~1. The linking number is half the sum of the assigned crossings $L(\mathcal{A},\mathcal{B})=\frac{1}{2}\sum_{i}\alpha_{i}(\mathcal{A},\mathcal{B})$, $\mathcal{A}$, and $\mathcal{B}$ are two UPOs. Self-linking number, on the other hand, is defined to be $L(\mathcal{A},\mathcal{A})=\sum_{i}\alpha_{i}(\mathcal{A},\mathcal{A})$. The linking Matrix will be of the form
\NiceMatrixOptions%
{code-for-first-row = \scriptstyle \text{$$},
code-for-first-col = \scriptstyle \text{$$}}
\begin{equation}
\mathrm{M}=
\begin{pNiceArray}{>{\strut}ccccc}[first-row,first-col=1,margin, extra-margin=2pt,colortbl-like]
& p_{1} & p_{2} & p_{3} & p_{4} & \cdots\\
p_{1} & SL_{11} & L_{12} & L_{13} & L_{14} & \cdots\\
p_{2} & L_{12} & SL_{22} & L_{23} & L_{24} & \cdots\\
p_{3} & L_{13} & L_{23} & SL_{33} & L_{34} & \cdots\\
p_{4} & L_{14} & L_{24} & L_{34} & SL_{44} & \vdots\\
\cdots & \vdots & \vdots & \vdots & \vdots & \ddots
\end{pNiceArray}
.
\end{equation}
\begin{figure}
\begin{center}
\includegraphics[width=2.5cm]{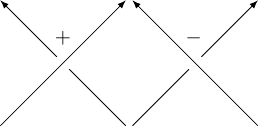}
\caption{$+,-$ are not affected by clockwise and counter-clockwise rotations.}
\end{center}
\end{figure}

While linking numbers are topological invariants, the self-linking numbers are not unless the underlying cyclic phase space is of type $\mathbb{R}^{2}\times S^{1}$.
\subsection{Global topological equivalence and chaotic perception}
While topological equivalence of dynamics can be local, having two strange attractors with the same linking matrix will refer to such equivalence globally. One possible usage of global equivalence could be in face recognition, in a narrow sense of course. After descrying the presence of chaos in biological neural networks\cite{Matsumoto:1987}, ``Which artificial networks are generative of chaos?'', ``What does chaos justify in real networks?'' have been among researchers' concerns. One suggested theory for this latter question is chaotic perception. Initially, in the late 80s, the functionality of discrimination was credited to limit cycles\cite{Skarda:1987}. Then, Ashwin and Timme argued that chaos might afford a means to discriminate different olfactory stimuli\cite{Ashwin:2005}. In their paper\cite{Marro:2007}, Marro and colleagues proposed a model for dynamic associative memory based on hopping from regularity to irregularity (chaos), the latter mimicking the brain states of attention and searching.
Suppose we are to recognize a person from a group of people. From experience, we know that the brain is so fast in doing so, a trait that is justified thanks to face symmetry (hence minimizing the information) and other distinguishing factors like gender, age, color, stature, and facial shape. Now, what if we are to distinguish identical twins? In case we do not have any of the mentioned features to resort to, how does a brain decide who is who? This is the place we may hypothesize that low-dimensional strange attractors may come to help the following way.\\
\textbf{Chaotic perception.} {\em Seeing objects of perception as invariant sets of fractal dimension, we may see the act of recognition as the comparison of the fingerprints of two globally equivalent strange attractors, one constructed and saved already when learning the object and the other one reconstructed at the time of perceiving the object.}\\
The topology of chaos teaches us that linking matrices are fingerprints of attractors with unique off-diagonal and non-unique diagonal entries. While the former entries may refer to the twins as a group, the latter would designate a matrix for each group member.
\section{Demands and challenges}
The concept of chaotic perception demands associating a neuronal model with the visual cortex first. In doing so, we need to note that one attributed functionality to chaos in neuroscience is ``storage and retrieval''\cite{Freeman:1992}. In infantile amnesia and Alzheimer's disease, one suffers from dementia - transparently, the small size of the network must be the culprit in engendering such suffering. Therefore, searching for a model with chaos in higher connectivity (larger size) would seem more realistic. One such neuronal model\cite{Sompolinsky:1988} is
\begin{equation}
\dot{X}_{i}(t)=-X_{i}(t)+\sigma\sum_{j=1}^{N}J_{ij}\phi(X_{j}(t)),
\end{equation}
comprised of $N$ fully connected neurons, each described by activation variable $X_{i},i=1,2\cdots,N$. $\sigma$ is the gain parameter, modulating the strength of connections. \linebreak$\phi(X)=\tanh(X)$, mimicking neurons' spikes, transfers activations into firing rates. The connection matrix feeds on Gaussian distribution with zero mean, variance $\ddfrac{1}{N}$, and correlation $[J_{ij},J_{ji}]=\ddfrac{\eta}{N}$, $\eta\in[-1,1]$. We take $\eta$ to be zero here (see Brunel's paper\cite{Marti:2018} for $\eta\neq0$).
\subsection{The flexibility issue, the first challenge}
According to the definition of chaotic perception, the model (1.1) must be generative of topologically equivalent strange sets globally. Checking for the capacity of random neural networks in generating so is what we are challenged by. To see the challenge, let us measure the flexibility of the neuronal model in imitating a well-known attractor like Lorenz; that is, we question if there is a weight matrix through which a neuron of the network is generative of the $x-$coordinate in
\begin{equation}
\begin{dcases}
\frac{dx}{dt}=\sigma(y-x),\\
\frac{dy}{dt}= x(r-z)-y,\\
\frac{dz}{dt}=xy-\beta z,
\end{dcases}
\end{equation}
In the following, we will see no matter we choose the theoretical or the inspectorial way, it is not feasible to know if there is a matrix, making our wish come true.

{\em The theoretical way.} In 2017, Pikovsky\cite{Pikovsky:2017} proposed his linear algebra approach to reconstruct the coupling matrix of (1.1) from an observed time series. We invert the model to have
\begin{equation}
\phi(X_{j})=\frac{1}{\sigma}\sum_{i}W_{ji}\underbrace{(\dot{X}_{i}+X_{i})}_{\equiv Y_{i}(t)},\quad W=J^{-1}.
\end{equation}
By looking for $M$ pairs of $(t_{1},t_{2})$ (which depends on the length of a series) such that $X_{j}(t_{1})\approx X_{j}(t_{2})$, then
\begin{equation}
\sum_{i}W_{ji}Z^{(m)}_{i}\approx0,
\end{equation}
where $m=1,2,\cdots, M$ and $Z_{i}=Y_{i}(t_{2})-Y_{i}(t_{1})$. We apply the singular value decomposition, $Z=U\sum V^{T}$, to have $W_{ji}=V_{N}$ ($N$ is the rank in fact, following from no correlation between the etriers). \\
With the augmentation of the network's size, which is required for chaos emergence, the calculation of the inverse matrix requires significant advances in the analysis of random matrix theory.\\
{\em The inspectorial way.} In this method, we resort to optimization tools. To get the same series, the function to be minimized must be defined in terms of linking matrices\footnote{We already know that non-fractal objects (spheres, cubes, tetrahedrons, etc.) could differ globally but have the same topological dimension. Likewise, fractal objects could differ globally but have the same fractal dimension. We may also have many globally-different chaotic attractors with the same metric entropy. Also, because convergence and divergence are conflictive arguments, it is no way to optimize all the exponents at once according to the Pareto set. Therefore, we must define the fitness function in terms of topological attributes.}. Clearly, the inspection through which the matrices are derived makes heuristic algorithms like ``Genetic''\cite{Kramer:2017} the apt tool to employ. One possible definition of the fitness function can be\footnote{Lor. stands for Lorenz.}
\begin{equation}
\mathrm{F}\equiv\|\mathrm{M}_{\text{net}}-\mathrm{M}_{\text{Lor.}}\|=\sum_{i=1}^{5}\sum_{j=1}^{5}|a_{ij}-b_{ij}|.
\end{equation}
To calculate the matrices, we need the embedded unstable periodic orbits. There are a couple of algorithms to extract the orbits; some requiring dynamical equations to accomplish the act of extraction. Because we must detect UPOs directly from chaotic time series, the algorithm presented by Wu et. al. \cite{Olyaei:2017} is by far easy to use than its precursors. Check appendix A to see how it works.

Inserting the numerical values tabulated in table 1,
\begin{table}
\caption{\label{tab:table2}Numerical values of the parameters for reconstructing the Lorenz attractor and calculating the linking matrix.}
\begin{tabular}{cc}
\hline
Input& (-0.95,-0.05,-0.05) \\
Length& 2500 points\\
$\delta t$ &0.01\\
$\sigma$& 10 \\
$\rho$& 28 \\
$\beta$& 8/3 \\
$\epsilon$& 0.01 \\
$k$ & 35\\
$p$ & 5\\
N& 250\\
Mean value&0\\
$\sigma$& 1.75\\
Variance& 1/250\\
Population size& 100\\
Number of iteration& 1000\\
Mutation rate& 0.1\\
\hline
\end{tabular}
\end{table}
fig.~2 indicates the reconstructed Lorenz attractor from the $x-$coordinate of the Lorenz system. The associated UPOs are given in fig.~3. We must note that $(P+1)-$period  orbit must be a continuation of $P-$period one. So, a LLL-RR or LL-RRR (L: left, R: right) 5-period orbit is required to take the second 4-period into consideration. The yielded linking matrix is
\NiceMatrixOptions%
{code-for-first-row = \scriptstyle \text{$$ },
code-for-first-col = \scriptstyle \text{$$ }}
\begin{equation}
\mathrm{M}_{\mathrm{Lor.}}=
\begin{pNiceMatrix}[first-row,first-col=1]
& 2 & 3 & 4 & 5\\
2 & 0 & 0 & 0 & 0\\
3 & 0 & 1 & 1 & 1\\
4 & 0 & 1 & 2 & 2\\
5 & 0 & 1 & 2 & 1
\end{pNiceMatrix}
.
\end{equation}
\begin{figure}[h]
    \centering
\includegraphics[width=0.25\textwidth]{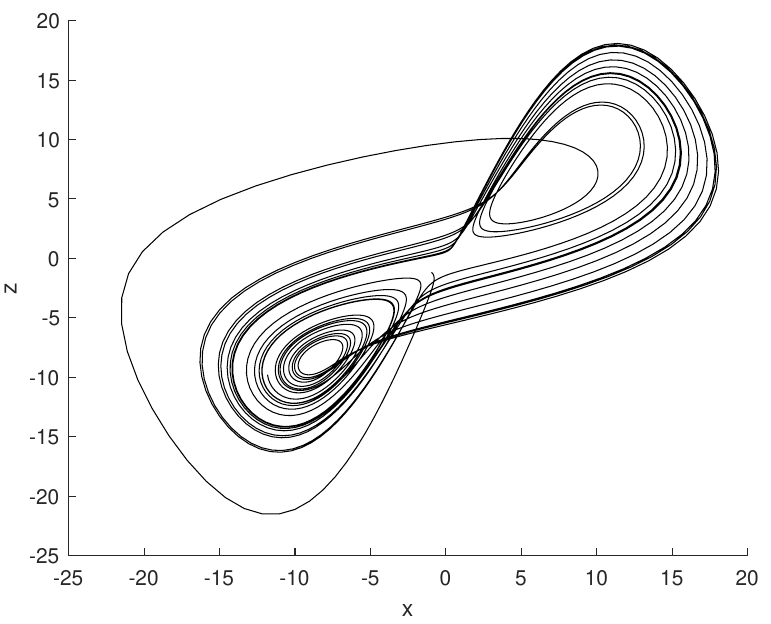}
\caption{Left: the reconstructed attractor from the $x-$coordinate of the Lorenz equation. This is the attractor we desire to extract through the network.}
\includegraphics[width=6cm]{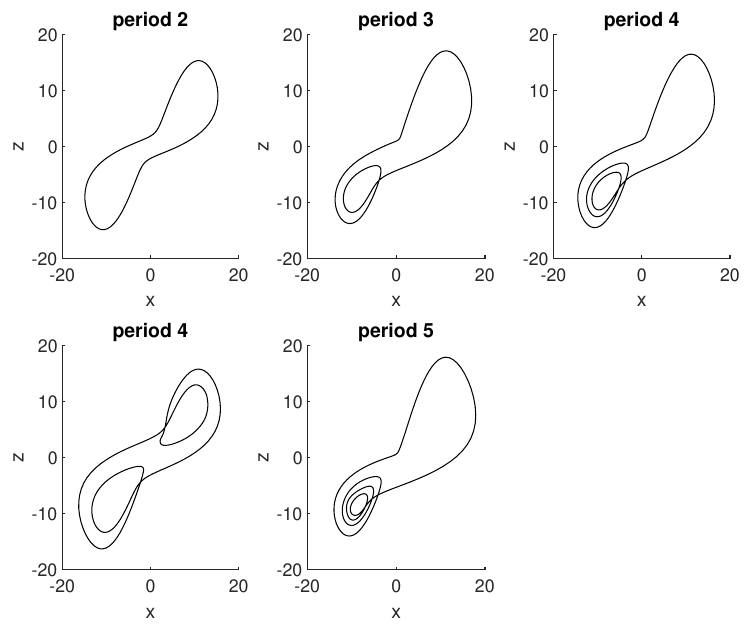}
\caption{(a) period 2 (Left-Right), (b) period 3 (LL-R), (c) period 4 (LLL-R,LL-RR), (d) period 5 (LLLL-R) unstable periodic orbits of reconstructed Lorenz attractor in fig.~2.}
\end{figure}

{\em Note: It is safe to use whatever we consider numerically for the UPO extraction of the well-known attractor when searching for the solution through the network.} \\
This follows from our observation when manipulating the lag, the sampling period, or the length of the network's outputs:
\begin{itemize}
\item \textit{Changing $\epsilon$.} \\
Reconstructions demand time lags, which require mutual information in turn. Let us recall the algorithm:

\hrulefill
\begin{small}
\begin{enumerate}
\item Generate a series of length L;
\item Set $\epsilon$ a value;
\item Calculate the probability of each entry, say $i$, in series by
\begin{equation*}
P(x_{i})=\frac{\text{number of $j$s such that $|x_{j}-x_{i}|<\epsilon$}}{L},\quad i,j=1,\cdots,L;
\end{equation*}
\item Set $\tau$ a value (we are going to sketch $I-\tau$ figure);
\item Create two-column Mutual Information Matrices (MIM)
\begin{equation*}
X_{i}=
\begin{pmatrix}
  x_{i} & x_{i+k} \\
 \end{pmatrix}
 ,\quad
 \begin{aligned}[t]
 i
 &= 1,\cdots,L-k,\\
 k
 &= 1,\cdots,\tau;
 \end{aligned}
\end{equation*}
\item Calculate the probability of each row in MIM through
\begin{equation*}
P(X_{i,k})=\frac{\text{number of $j$s such that $|X_{j}-X_{i}|<\epsilon$}}{L-k},\quad i,j=1,\cdots,L-k;
\end{equation*}
\item Use
\begin{equation*}
I(k)= P(X_{i,k})\log_{2}\Biggl(\frac{P(X_{i})}{P(x_{i})P(x_{i+k})}\Biggr)
\end{equation*}
to calculate the mutual information;
\item Sketch $I-\tau$ and find the first minimum.
\end{enumerate}
\end{small}
\hrulefill\\
\textnormal
The attractor and unstable periodic orbits we get if we change $\epsilon$ to $0.001$ are depicted in figs~.~4 and 5. In this case, we derive
\begin{equation}
\mathrm{M}_{\mathrm{Lor.}}=
\begin{pNiceMatrix}
0 & 1 & 0 & 0\\
1 & 1 & 1 & 3\\
0 & 1 & 2 & 0\\
0 & 3 & 0 & 1
\end{pNiceMatrix}
.
\end{equation}
for the linking matrix.

\begin{figure}[h]
\begin{center}
\includegraphics[width=2.5cm]{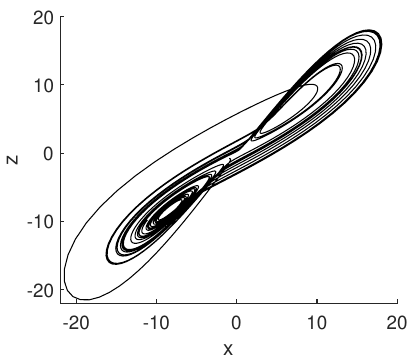}
\caption{The attractor derived from the same numerical values tabulated in the table 1 except $\epsilon=0.001$. Changing the lag of an attractor may alter the linking matrix.}
\includegraphics[width=6cm]{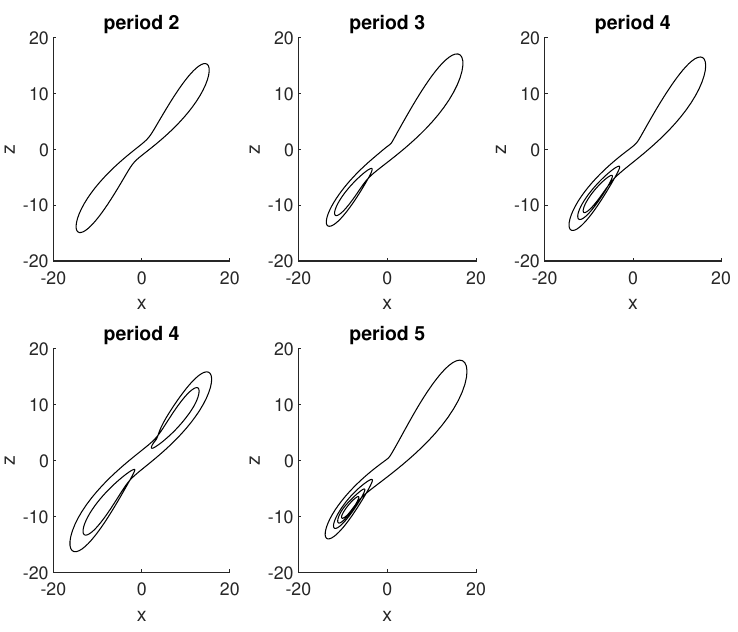}
\caption{The unstable periodic orbits associated with the attractor in fig.~4.}
\end{center}
\end{figure}
\item \textit{Changing the sampling period $\delta t$.}
Surly, we are not allowed to put any number for $\delta t$. If you take it to be $0.011$, then we have the attractor in fig.~6(left) for which we do not get any periodic orbit. For a legitimate $\delta t=0.0106$, the attractor and unstable periodic orbits we get are depicted in figs.6 (right) and 7. In this case, we derive
\begin{equation}
\mathrm{M}_{\mathrm{Lor.}}=
\begin{pmatrix}
0 & -1 & 1 & 0\\
-1 & -1 & -1 & 0\\
1 & -1 & 0 & -2\\
0 & 0 & -2 & 3
\end{pmatrix}
\end{equation}
for the linking matrix.
\begin{figure}[h]
    \centering
    \subfloat{%
        \includegraphics[width=0.2\textwidth]{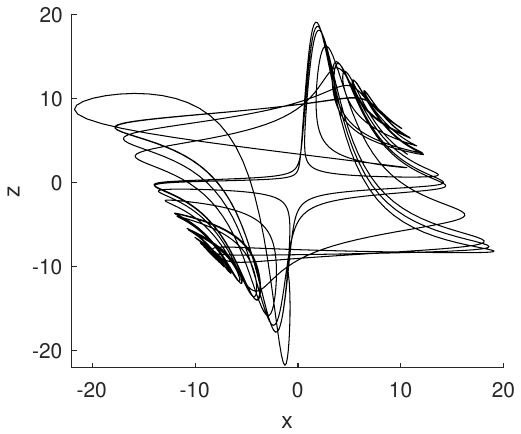}%
        \label{fig:a}%
        }%
    \subfloat{%
        \includegraphics[width=0.2\textwidth]{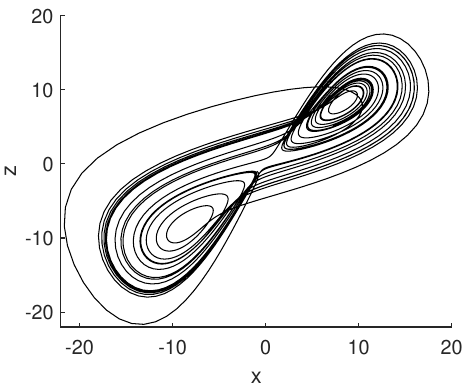}%
        \label{fig:b}%
        }%
    \caption{Left: the attractor derived from the same numerical values tabulated in the table 1 except $\delta t=0.011$. We do't get any unstable period orbit for this attractor. Right: the attractor derived from the same numerical values tabulated in the table 1 except $\delta t=0.0106$. Changing the steps in a time series may alter the linking matrix.}
\includegraphics[width=8cm]{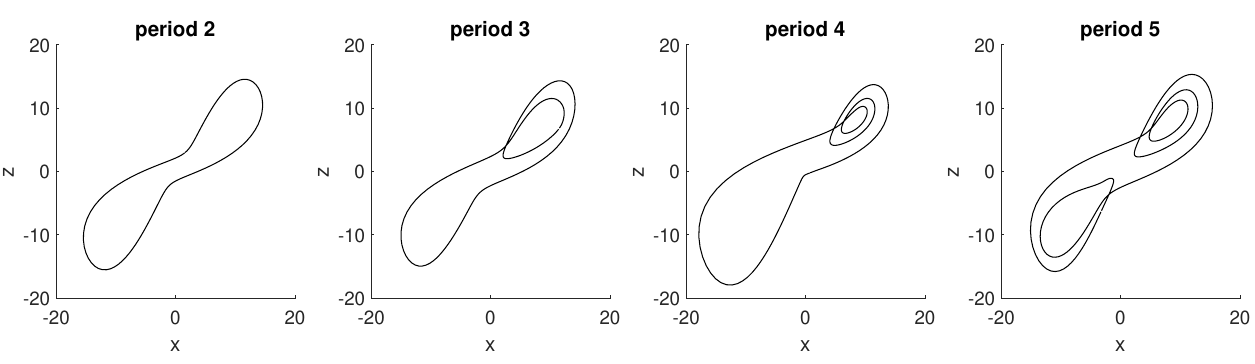}
\caption{The unstable periodic orbits associated with the attractor in fig.~6, right.}
\end{figure}
\item \textit{Changing the length.}
If we increase the length of the series to some higher number say $5000$, the attractor and unstable periodic orbits we get are depicted in figs~.8 and 9. In this case, we derive
\begin{equation}
\mathrm{M}_{\mathrm{Lor.}}=
\begin{pNiceMatrix}
0 & 0 & 0 & 0\\
0 & 1 & 1 & 2\\
0 & 1 & 2 & 2\\
0 & 2 & 2 & 3
\end{pNiceMatrix}
\end{equation}
for the linking matrix.
\begin{figure}[h]
\begin{center}
\includegraphics[width=2.5cm]{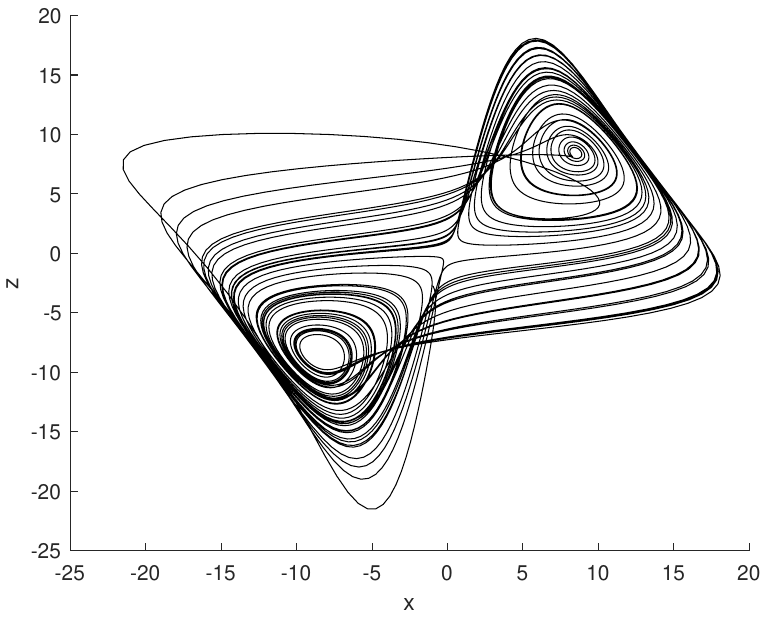}
\caption{The attractor driven from the continued series in fig.~2 to the length 5000. Changing the length of a time series may alter the linking matrix.}
\includegraphics[width=8cm]{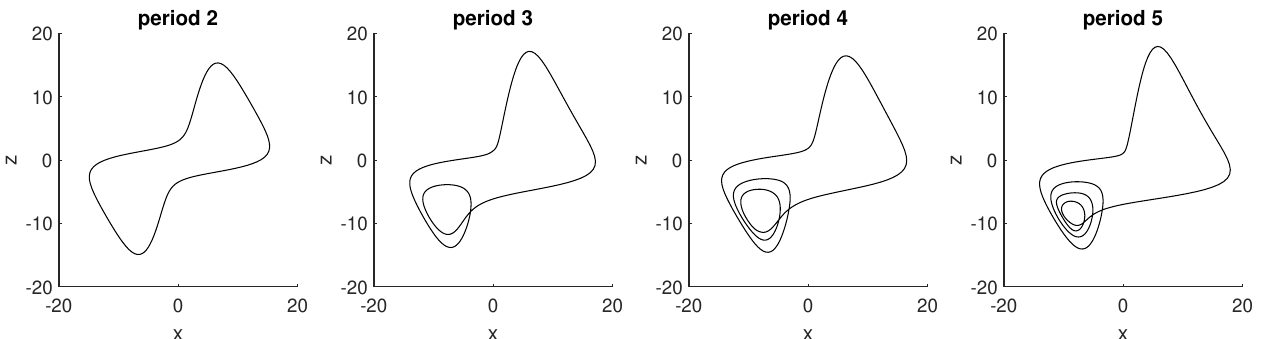}
\caption{The unstable periodic orbits associated with the attractor in fig.~8.}
\end{center}
\end{figure}
\end{itemize}

Accordingly, the attractors in figs.~2, 4, 6, and 8 cannot be considered as the same.

Finally comes the steps to calculate the linking number of the network's series:
\begin{enumerate}
\item Making the last neuron of the network responsible for the construction and reconstruction of globally equivalent attractors, generate the series through a population of weight matrices, and do the reconstruction.
\item Filtering I: rapidity is a looked-for feature in chaotic perception and hyperchaos would be inimical to this end for sure. Besides, the authenticity of the topological invariants just in 3-space is another factor to put an attractor of embedding dimension\cite{Cao:1997} (the minimum dimension in which an attractor does not suffer from self-intersections) four and higher aside and pick out low-dimensional series to work with;
\item Filtering II: detect the notches (Appendix A) in $\overline{\xi}(\omega_{k})$ germane to the series left in the preceding step and discard the ones with notches less than 4;
\item Filtering III: sort the periodic orbits based on the number of intersections (in the Lorenz case, 0, 1, 2, and 3 respectively);
\item Filtering IV: keep merely the series with ``$\omega_{0}$ of period $P$ $>$ $\omega_{0}$ of period $P+1$ $>$ $\omega_{0}$ of period $P+2$ $>$ $\omega_{0}$ of period $P+3$'';
\item Calculate the linking matrix associated with the remaining series and choose the minimum one.
\end{enumerate}
Seemingly, the inspectorial way does not pose the problem the theoretical way did. As a matter of fact, it does; the challenge is hidden in step (1). Given the series $x_{1},x_{2},\cdots,x_{N}$ with embedding dimension 3 and time lag $\tau$, the $x,y,z$ coordinates of the reconstructed attractor are
\begin{equation}
\begin{pNiceMatrix}[first-row=1]
x & y & z\\
x_{1} & x_{\tau+1} & x_{2\tau+1}\\
x_{2} & x_{\tau+2} & x_{2\tau+2}\\
\vdots & \vdots & \vdots\\
x_{L-2\tau} & x_{L-\tau} & x_{L}
\end{pNiceMatrix}
,
\end{equation}
$L$ being the length of the series. $(x_{1},x_{\tau+1},x_{2\tau+1})$, the initial point, must reside in the basin of attraction of the fixed point(s) embedded in the attractor we are chasing after. Therefore, the problem reduces to see if there are fixed point(s) of the neuronal model distributed (in isolation) the way they are along with their (in)stability. Knowing this requires a knowledge of the distribution of the fixed points in $\mathbb{R}^3$, which are nothing but the projection of the fixed points (of the neuronal model) living in $\mathbb{R}^n$. In 2013, G.~Wainreb, et. al. showed that in transition from fixed point to chaotic zone, the only fixed point of the net grows exponentially\cite{Touboul:2013} as
\begin{equation}
\mathbb{E}(A_n(\sigma))\sim e^{n(\sigma - 1)^2},\quad n\to \infty
\end{equation}
where $\mathbb{E}$ is the expectation value of $A_n$, the number of points satisfying $X_{i}=\sigma\sum_{j=1}^{N}J_{ij}\phi(X_{j})$. As with the theoretical way, realizing the exact distribution of fixed points requires advances in random matrix theory. We need to note that {\em having the distribution will not guarantee the existence of flexibility.}

Unfortunately, for more than three decades the flexibility of random neural networks in practice has been overlooked; it's been targeted by researchers\cite{Ravid:2024} recently. It is not just the perception that makes the flexibility issue important. If strange attractors are detected thanks to the collaboration of units in a real network (social, biological, ecological, physical, chemical, etc.), what mathematical model will be attributable to the dynamics of such assemblies? Surely, the one that can yield the same invariant sets. Accordingly, we expect the mathematical models to be imitative beyond mere generative.

\subsection{Chaos preserving matrices, the second challenge}
The concept of chaotic perception requires speaking to the object invariance, an interest of theoretical neuroscientists currently\cite{Sompolinsky:2018}. If side and frontal views (in fact, we can take any two perspectives of perception) of an object amount to two different time series, the object must be associated with one strange attractor to respect object invariance. Two series mean two different weight matrices, giving two attractors with the same linking numbers. Vividly, we must get one of the attractors by deforming the other, following from one of the topics in chaos topology - deforming an attractor so that its topological invariants remain unchanged. The second challenge is ``how'' we can check the connection between matrices yielding the same attractor. Are they infinitesimal deviations from each other? Because if yes, the act of perception rather than the phase space, as implied by the definition of the chaotic perception, would take place in the matrix space to which the connectivity matrix of the model belongs. Mathematically, using the Euler's method, we are looking for the relation between $W,W'$ in
\begin{equation}
\begin{aligned}[t]
\underbrace{
\begin{pmatrix}
w_{11} & w_{12} & \cdots & w_{1n}\\
w_{21} & w_{22} & \cdots & w_{2n}\\
\vdots & \vdots & \ddots & \vdots \\
w_{n1} & w_{n2} & \cdots & w_{nn}
\end{pmatrix}
}_{W}
\underbrace{
\begin{pmatrix}
\phi(x_{1}(t))\\
\phi(x_{2}(t))\\
\cdots\\
\phi(x_{n}(t))
\end{pmatrix}
}_{X}
&=
\underbrace{
\begin{pmatrix}
\frac{x{1}(t+1)-x_{1}(t)}{s}+x_{1}(t)\\
\frac{x{2}(t+1)-x_{2}(t)}{s}+x_{2}(t)\\
\vdots \\
\frac{x{n}(t+1)-x_{n}(t)}{s}+x_{n}(t)
\end{pmatrix}
}_{Y}
, t=0,s,2s,\cdots\\
W'X'
&=Y'
\end{aligned}
\end{equation}
such that
\begin{equation}
\begin{pNiceMatrix}
x_{n}(0) & x_n(\tau) & x_n(2\tau)\\
x_{n}(s) & x_n(\tau+s) & x_n(2\tau+s)\\
\vdots & \vdots & \vdots\\
x_{n}(L-2\tau) & x_n(L-\tau) & x_n(L)
\end{pNiceMatrix}
,
\begin{pNiceMatrix}
x'_{n}(0) & x'_n(\tau) & x'_n(2\tau)\\
x'_{n}(s) & x'_n(\tau+s) & x'_n(2\tau+s)\\
\vdots & \vdots & \vdots\\
x'_{n}(L-2\tau) & x'(L-\tau) & x'_n(L)
\end{pNiceMatrix}
\end{equation}
have the same linking matrices. The lag may also change, but not to the point it changes the fingerprint (the case in Fig.~4).

\appendix
\section{The extraction of unstable periodic orbits}
In short, assume that chaotic $x$ has an unstable periodic orbit with the minimal period $T$. We may express $x$ through its Fourier series
\begin{equation}
x=\sum_{n=-k}^{k}a_{n}e^{in\omega t},\quad i=\sqrt{-1},\omega=\frac{2\pi}{T}.
\end{equation}
The same periodic motion can be due to a system of harmonic oscillators
\begin{equation}
\dot{A}_{n}=in\omega A_{n},\quad n=-k,\cdots,k,
\end{equation}
where dot is differentiation with respect to the time. If $A_{n}=a_{n}e^{in\omega t}$, then \linebreak$x-\sum_{n=-k}^{k}A_{n}$ represents the deviation of $x$ from its periodic behaviour. We bridge two systems through a coupling $K_{n}$ such that
\begin{equation}
\dot{A}_{n}=in\omega A_{n}+K_{n}\Biggl(x-\sum_{n=-k}^{k}A_{n}\Biggr),
\end{equation}
$K_{-n}$ being the complex conjugate of $K_{n}$. Eq. (13) can be written as
\begin{equation}
\dot{A}=\Omega A+\kappa x,
\end{equation}
where
\begin{equation}
\begin{split}
&A=(A_{k},A_{k-1},\cdots,A_{-k+1},A_{-k})^{\mathrm{T}}\\
&\kappa=(\kappa_{k},\kappa_{k-1},\cdots,\kappa_{-k+1},\kappa_{-k})^{\mathrm{T}}\\
&\Omega=
\begin{pmatrix}
i\omega-\kappa & -\kappa & -\kappa & \dots & -\kappa\\
-\kappa & i2\omega-\kappa & -\kappa & \dots & -\kappa\\
\vdots & \vdots & \vdots & \cdots & \vdots\\
-\kappa & -\kappa & -\kappa & \cdots & iN\omega-\kappa
\end{pmatrix}
.
\end{split}
\end{equation}
$\mathrm{T}$ denotes the transpose of a matrix. Using Euler's exponential integrator, (3.6) turns into
\begin{equation}
A(t_{j+1})=e^{\omega\delta t}A(t_{j})+\frac{e^{\Omega\delta t}-I}{\Omega}\,\kappa x(t_{i}).
\end{equation}
The components of the unstable periodic orbit are given by
\begin{equation}
x_{n}=\frac{1}{L_{0}}\sum_{i=1}^{L_{0}}A_{n}(\tau_{0}+l\delta t)e^{-in\omega_{0}(\tau_{0}+l\delta t)},
\end{equation}
$L_{0}$ being the number of samples in the interval $\tau_{0}\leqslant t\leqslant \tau_{0}+\frac{2\pi}{\omega_{0}}$. $\omega_{0}$, and $\tau_{0}$ for different periods (p) are derived respectively from the notches of
\begin{equation}
\overline{\xi}(\omega_{k})=\frac{\sum_{i=1}^{J-L_{k}}\langle \xi(\tau_{i},\omega_{k})\rangle^{-p+2}}{\sum_{i=1}^{J-L_{k}}\langle \xi(\tau_{i},\omega_{k})\rangle^{-p + 1}},\quad \omega_{k}=\frac{2\pi}{L_{k}\Delta t}
\end{equation}
and
\begin{equation}
\langle \xi(\tau_{i},\omega_{k})\rangle=\sum_{n=-k}^{k}\Biggl|\frac{A_{n}(\tau_{j}+2\pi/\omega_{k})-A_{n}(\tau_{j})}{2\pi \kappa_{n}/\omega_{k}}\Biggr|^2,
\end{equation}

\bibliographystyle{amsplain}

\providecommand{\bysame}{\leavevmode\hbox to3em{\hrulefill}\thinspace}
\providecommand{\MR}{\relax\ifhmode\unskip\space\fi MR }
\providecommand{\MRhref}[2]{%
  \href{http://www.ams.org/mathscinet-getitem?mr=#1}{#2}
}
\providecommand{\href}[2]{#2}
\begin{thebibliography}{}

\end{thebibliography}


\begin{thebibliography}{10}
\bibitem {Kuznetsov:2004} Y. Kuznetsov, \textit{Elements of Applied Bifurcation Theory} Springer
(2004), ed.~3.
\bibitem {Gilmore:2011} R. Gilmore and M. Lefranc, \textit{The topology of chaos} Wiely (2011), ed.~2.
\bibitem {Hao:1989} B. Hao, \textit{Elementary symbolic dynamics and chaos in dissipative systems} World Scientific (1989), 343-347.
\bibitem {Matsumoto:1987} G. Matsumoto, and K. Aihara, and Y. Hanyu, and N. Takahashi, and S. Yoshizawa, and J. Nagumo, \textit{Chaos and phase locking in normal squid axons} Phys Lett A
\textbf{123} (1987), 162--166. {https://doi.org/10.1016/0375-9601(87)90696-7}
\bibitem {Skarda:1987} C.~A. Skarda and W.~J. Freeman, \textit{How brains make chaos in order to make sense of the world} Behav. Brain Sci
\textbf{10} (1987), 161--195.
\bibitem {Ashwin:2005} P. Ashwin and M. Timme, \textit{When instability makes sense} Nature
\textbf{436} (2005), 36--37.
\bibitem {Marro:2007} J. Marro, and J. Torres, and J. Cort\'{\i}s, \textit{Chaotic hopping between attractors in neural networks} Neural Networks
\textbf{20} (2007), 230--235.
\bibitem {Freeman:1992} W.~J. Freeman, \textit{Tutorial On Neurobiology: From Single Neurons to Brain Chaos} International Journal of Bifurcation and Chaos
\textbf{2} (1992), 451--482.
\bibitem {Sompolinsky:1988} H. Sompolinsky, and A. Crisanti, and H.~J. Sommers, \textit{Chaos in Random Neural Networks} Phys Rev Lett
\textbf{61} (1988), {https://doi.org/10.1103/PhysRevLett.61.259}.
\bibitem {Marti:2018} D. Mart\'{\i}, and N. Brunel, and S. Ostojic, \textit{Correlations between synapses in pairs of neurons slow down dynamics in randomly connected neural networks} Phys Rev E
\textbf{97} (2018), {https://doi.org/10.1103/PhysRevE.97.062314}.
\bibitem {Pikovsky:2017} A. Pikovsky, \textit{Reconstruction of a scalar voltage-based neural field network from observed time series} epl
\textbf{119} (2017), {https://doi.org/10.1209/0295-5075/119/30004}.
\textbf{110} (2013), {https://doi.org/10.1103/PhysRevLett.110.118101}.
\bibitem {Kramer:2017} O. Kramer, \textit{Genetic Algorithm Essentials} Springer (2017).
\bibitem {Olyaei:2017} A.~A. Olyaei, and C. Wu, and W. Kinser, \textit{Detecting unstable periodic orbits in chaotic time series using synchronization} Phys Rev E
\textbf{96} (2017),\\{https://doi.org/10.1103/PhysRevE.96.012207}.
\bibitem {Cao:1997} Liangyue Cao, \textit{Practical method for determining the minimum embedding dimension of a scalar time series} Physica D: Nonlinear Phenomena
\textbf{110} (1997), \\{https://doi.org/10.1016/S0167-2789(97)00118-8}.
\bibitem {Touboul:2013} G. Wainrib and Jonathan Touboul, \textit{Topological and Dynamical Complexity of Random Neural Networks} Phys Rev Lett
\bibitem {Ravid:2024} Ravid Shwartz-Ziv, \textit{Just How Flexible are Neural Networks in Practice?} (2024), arXiv:2406.11463v1.
\bibitem {Sompolinsky:2018} S. Chung, and D.~D. Lee, and H. Sompolinsky, \textit{Classification and Geometry of General Perceptual Manifolds} Phys Rev X
\textbf{8} (2018), {https://doi.org/10.1103/PhysRevX.8.031003}.
\end{thebibliography}

\end{document}